# Caesium fallout in Tokyo on 15th March, 2011 is dominated by highly radioactive, caesium-rich microparticles

Satoshi Utsunomiya [1], Genki Furuki [1], Asumi Ochiai [1], Shinya Yamasaki [2], Kenji Nanba [3], Bernd Grambow [4], Rodney C. Ewing [5],

[1]Department of Chemistry, Kyushu University, Motooka 744, Nishi-ku, Fukuoka 819-0395, Japan
[2]Faculty of Pure and Applied Sciences and Center for Research in Isotopes and Environmental Dynamics, University of Tsukuba, 1-1-1 Tennodai, Tsukuba, Ibaraki 305-8577 Japan
[3]Department of Environmental Management, Faculty of Symbiotic System Science, Fukushima University, Kanayagawa 1, Fukushima, 960-1296 Japan
[4] SUBATECH, IMT Atlantique, University of Nantes and CNRS/IN2P3, Nantes 44307, France
[5]Department of Geological Sciences and Center for International Security and Cooperation, Stanford University, Stanford, CA 94305-2115 USA

## ABSTRACT

In order to understand the chemical properties and environmental impacts of low-solubility Cs-rich microparticles (CsMPs) derived from the FDNPP, the CsMPs collected from Tokyo were investigated at the atomic scale using high-resolution transmission electron microscopy (HRTEM) and dissolution experiments were performed on the air filters. Remarkably, CsMPs 0.58–2.0 μm in size constituted 80%–89% of the total Cs radioactivity during the initial fallout events on 15th March, 2011. The CsMPs from Tokyo and Fukushima exhibit the same texture at the nanoscale: aggregates of Zn–Fe-oxide nanoparticles embedded in amorphous $SiO_2$ glass. The Cs is associated with Zn–Fe-oxide nanoparticles or in the form of nanoscale inclusions of intrinsic Cs species, rather than dissolved in the $SiO_2$ matrix. The Cs concentration in CsMPs from Tokyo (0.55–10.9 wt%) is generally less than that in particles from Fukushima (8.5–12.9 wt%). The radioactivity per unit mass of CsMPs from Tokyo is still as high as ~$10^{11}$ Bq/g, which is extremely high for particles originating from nuclear accidents. Thus, inhalation of the low-solubility CsMPs would result in a high localized energy deposition by β (0.51–12×$10^{-3}$ Gy/h within the 100-μm-thick water layer on the CsMP surface) and may have longer-term effects compared with those predicted for soluble Cs-species.

## 1. Introduction

In March 2011, ~$10^{17}$ becquerels (Bq) of radionuclides, including ~$10^{16}$ Bq of radioactive $^{134}$Cs and $^{137}$Cs, were released from the Fukushima Daiichi Nuclear Power Plant (FDNPP), Japan, owing to failure of the reactors[1–3]. Because of their relatively long half-lives, 2.065 and 30.17 years for $^{134}$Cs and $^{137}$Cs, respectively, these isotopes are currently major contributors to the radiation dose at the surface where they have accumulated. 1.2%–6.6% of the total $^{137}$Cs inventory have been released from the damaged reactors[4] and 22% of the released $^{137}$Cs was dispersed to the terrestrial environment in 15 prefectures across Japan, including the capital city Tokyo[5].

Previously, Cs was thought to have been released as soluble species: ~90% as CsOH and ~10% as CsI inferred from the previous experimental studies[6]. These soluble species were wet-deposited by precipitation[5,7,8] and immediately adsorbed onto clays at very low concentrations, <$10^3$ Bq/g[9,10]. The high affinity of Cs to clay minerals resulted in Cs immobilization in surface soils[9,11], causing the current high levels of gamma and beta radiation in the environment. In contrast, several recent studies reported Cs-rich microparticles (CsMPs) in surface environment near the FDNPP[12,13,14] and even approximately 170 km southwest of the FDNPP[15,16]. These CsMPs can potentially provide important insights into the reactions during the meltdowns[17] and another route for Cs migration into the environment; however, the nature of CsMPs and their impact on megacities have never been discussed previously. In Tokyo, the largest radioactive plume was detected at Setagaya ward on 15th March, 2011 (Fig. S1). The maximum radioactivities of $^{132}$Te, $^{131}$I, and $^{132}$I, $^{134}$Cs and $^{137}$Cs were measured to be 395, 241, 281, 64 and 60 Bq/m$^3$ during 10:00-11:00 on 15th March, 2011[18]. Although the previous study analyzed the bulk radioactivities on the filters, the chemical form of the radionuclides, especially Cs as a form of CsMP, has never been determined. Thus, we undertake a quantitative analysis of radioactivity from CsMPs from dissolution experiments on atmospheric particles collected in Tokyo. The chemical and structural properties of CsMPs collected from Tokyo are examined at the atomic scale and compared with a sample collected from Fukushima using high-resolution transmission electron microscopy (HRTEM).

## 2. Methods

The aerosol sample was collected 1 m above the ground in Setagaya Ward, Tokyo, which is ~230 km southwest of the FDNPP. The sampling was conducted at 0.6 m$^3$/min using a high-volume air sampler (Staplex® model TFIA) with a glass fibre filter (ADVANTEC GD-120R). Air filter #1 and #2 were sampled during 0:00–7:00 and 15:00–16:00 JST on 15th March, 2011. The details of the sampling procedure and the bulk radioactivity data were described in a previous study[18].

The soil sample was collected from the top ~1 cm of paddy soil in Ottozawa, Okuma Town, Futaba County, Fukushima, on 16th March, 2012. The soil was mainly composed of clay minerals, quartz and feldspars. As entry to the area is still restricted prohibited owing to the high radiation dose, the locality had not been artificially disturbed. The radiation dose ~1 m above the ground was 84 µSv/h.

*Corresponding author
E-mail: utsunomiya.satoshi.998@m.kyushu-u.ac.jp

The separation procedure of CsMP from the air filter is illustrated in Fig. S3. Briefly, the filter was cut into nine small pieces and placed on double-stick carbon tape regularly arrayed on the grid paper. Tiny pieces were cut from the filter paper, which was then covered with plastic wrap and exposed to an imaging plate (Fuji film, BAS-SR 2025) for 1-12 h in a dark box. The autoradiograph image was obtained in digital format at a resolution of 25 µm using an IP reader (GE, Typhoon FLA9500). A few pieces of the filters with intense spots in the autoradiograph image were placed into 0.1 mL of ultra-pure water in a microtube and sonicated for 10 min to dissociate the CsMPs from the filter. Droplets of the solution were arrayed on a glass slide and allowed to dry completely. The dried spots were contacted with double-stick carbon tape, and the tape was again cut into small pieces. Subsequently, the piece of carbon tape with the CsMP was identified from the autoradiograph images. The piece was placed on an aluminium plate and coated with C using a carbon coater (SANYU SC-701C) prior to scanning electron microscopy (SEM) analysis. The CsMP was searched for using an SEM (Shimadzu, SS550 and Hitachi, SU6600) equipped with energy dispersive X-ray spectrometry (EDX, EDAX Genesis). The acceleration voltage was 5-25 kV for imaging details of the surface morphology and 15-25 kV for elemental analysis, including area analysis and elemental mapping. Separation of CsMPs from soil samples was proceeded based on our previous study[17].

The FIB (FEI, Quanta 3D FEG 200i Dual Beam) was utilized to prepare a thin foil of individual CsMPs with diameters of a few µm. The ion source was Ga, while W deposition was used to minimize damage by the ion bombardment. The current and accelerating voltage of the ion beam were adjusted to be 100 pA to 30 nA and 5–30 kV depending on the progress of thinning and sample properties such as hardness and size. The thinned piece was attached to the semilunar-shaped Cu grid for FIB and further thinned by an ion beam at 5 kV.

HRTEM, with EDX and high-angle annular dark-field scanning transmission electron microscopy were performed using JEOL JEM-ARM200F and JEM-ARM200CF with an acceleration voltage of 200 kV. The JEOL Analysis Station software was used to control the STEM-EDX mapping. To minimize the effect of sample drift, a drift-correction mode was used during acquisition of the elemental map. The STEM probe size was ~0.13 nm, generating ~140 pA of current when 40 µm of the condenser lens aperture was inserted. The collection angle of the HAADF detector was ~97–256 mrad.

The $^{134}$Cs and $^{137}$Cs radioactivities of the CsMPs were determined by gamma spectrometry. The radioactivity of the additional microparticle, which was only ~400 µm in size, from surface soil in Fukushima was precisely determined at the radioisotope centre in Tsukuba University, Japan, and utilized as a standard point specimen for $^{134}$Cs and $^{137}$Cs. The radioactivity of the point source standard was 23.9 Bq for $^{134}$Cs and 94.6 Bq for $^{137}$Cs as of 29 September 2015. Measurement of radioactivity was performed on CsMPs as well as the point source standard using a germanium semi-conductor detector, GMX23 (SEIKO E&G), GMX30 (SEIKO E&G), GMX40 (SEIKO E&G) and GX6020 (Canberra) at the centre for radioisotopes in Kyushu University, Japan. The acquisition time was 434428 s for TKY6 using GMX23, 160447 s for TKY1 using GMX30, 89252 s for TKY2 and 91040 s for TKY3 and 262570 s for TKY4 and 322643 s for TKY5 and 86414 s for OTZ using GMX40, 616471.7 s for TKY7 using GX6020 .

Energy deposition, E0-100 (Gy/h), of β radiation in water within the spherical space between the surface of the CsMP and 100 µm from the surface can be approximately calculated based on the following equation:

$$E_{0-100} = \frac{LET \times R \times 100[\mu m] \times 1.6022 \times 10^{-16}[J/eV] \times 3600[s]}{M_w} \quad (1)$$

*R*: Radioactivity of β radiation (Bq) estimated based on γ radioactivity

*LET*: Linear energy transfer is defined as the energy transferred to the target material while a single β particle travels a unit length (keV/µm). The *LET* values are assumed to be 0.40 keV/µm for β rays in the present calculation.

$M_w$ : Mass of water (kg) within 100 µm of the particle's surface, which can be calculated by

$$M_w = \left\{\left[\frac{4}{3}\pi \times (r_p + 100)^3\right] - \left[\frac{4}{3}\pi \times r_p^3\right]\right\} \times 10^{-15} \quad (2)$$

where $r_p$ is the particle radius.

Energy deposition of γ radiation in water is negligible in the present calculation. The number of molecules produced by radiolysis, $N_{rad}$ (molecules/s), in water within the spherical space between the surface of the CsMP and 100 µm from the surface (Fig. S4) can be calculated as follows:

$$N_{rad} = \frac{E_{0-100} \times M_w \times Y_i \times 10^{-6}}{3600} \times N_A \quad (3)$$

$Y_i$: Product yield (µmol/J) [30] of chemical species *i*

$N_A$: Avogadro's number

Dissolution experiments were conducted to evaluate the contribution of CsMP to total Cs radioactivity. A piece of air filter was immersed in ultra-pure water for 24 h at room temperature statically such that the CsMPs trapped on the filter would not become suspended in the solution. After the immersion experiment, the filter was carefully removed from the pure water and dried in air. Autoradiography and gamma spectrometry were performed on the filter before and after the dissolution experiment. The solution after the dissolution experiment was filtered using a 0.1-µm membrane filter and the radioactivity was analysed by gamma spectrometry. The autoradiograph made prior to the experiment contained almost the same distribution of hot spots and did not exhibit any significant difference from the autoradiograph obtained after the dissolution experiment.

## 3. Results

Atmospheric particles were collected at Setagaya ward, Tokyo[18]. Air filters #1 and #2 were sampled during 0:00–7:00 and 15:00–16:00 JST on 15th March, 2011, respectively. The optical images and autoradiographs of these two filters show the presence of dust as well as small but numerous radioactive spots on the filters (Fig. 1a and b). Seven CsMPs were successfully separated from air filter #2; these are hereafter labelled as TKY1–7. For comparison, a single CsMP was also separated from surface soil of a paddy field in Ottozawa, Okuma

**Table 1:** Summary of the radioactivity and associated parameters of the TKY and OTZ CsMP samples.
*The radioactivity was decay-corrected to 15:36 JST on 12th March, 2011,.
**The radioactivity per unit mass was calculated assuming that the particles are spherical in shape, with a density of 2.6 g/cm$^3$.
***Data from Furuki et al. (2016)[17].

| Sample | Particle size (μm) | Radioactivity (Bq)* | | $^{134}Cs/^{137}Cs$ | Radioactivity per unit mass (Bq/g)** | Cs concentration as $Cs_2O$ (wt%) determined by EDX analysis | |
| --- | --- | --- | --- | --- | --- | --- | --- |
| | | $^{134}Cs$ | $^{137}Cs$ | | | By SEM-EDX | By TEM-EDX |
| **TKY1** | 1.7 | 1.19(±0.038) | 1.09(±0.014) | 1.09 | 3.42×10$^{11}$** | 7.77 | n/a |
| **TKY2** | 1.4 | 0.403(±0.114) | 0.426(±0.011) | 0.948 | 2.04×10$^{11}$** | 3.13 | n/a |
| **TKY3** | 2.0 | 0.516(±0.038) | 0.522(±0.012) | 0.989 | 9.52×10$^{10}$** | 2.07 | 0.81-2.1 |
| **TKY4** | 1.0 | 0.273(±0.020) | 0.261(±0.006) | 1.04 | 3.93×10$^{11}$** | 6.86 | 5.8-10.9 |
| **TKY5** | 1.2 | 0.228(±0.019) | 0.207(±0.005) | 1.10 | 1.85×10$^{11}$** | 3.24 | 0.55-8.2 |
| **TKY6** | 1.0 | 0.243(±0.020) | 0.258(±0.006) | 0.942 | 3.68×10$^{11}$** | 3.83 | n/a |
| **TKY7** | 0.58 | 0.0453(±0.016) | 0.0484(±0.012) | 0.936 | 3.53×10$^{11}$** | 7.19 | n/a |
| **OTZ*** | 2.0 | 2.00(±0.080) | 2.07(±0.031) | 0.967 | 3.74×10$^{11}$** | 8.32 | 8.50-12.9 |

town in Fukushima, ~3.9 km west of the FDNPP[9]: this particle is labelled as OTZ.

*Radioactive properties of the TKY CsMPs*

Table 1 summarizes the radioactivity of the CsMPs and the associated parameters. The $^{134}Cs/^{137}Cs$ isotopic ratio is 0.94–1.1, average 1.01, which confirms that these CsMPs were originated from FDNPP. The radioactivity per unit mass calculated assuming the density of $SiO_2$ glass to be 2.6 g/cm$^3$ is between $9.5 \times 10^{10}$ and $3.9 \times 10^{11}$ Bq/g.

*Nano-structure and chemical properties of TKY CsMPs*

Most of the TKY CsMPs are round in shape and 0.58–2.0 μm in size (Fig. 2). The scanning transmission electron microscopy energy dispersive X-ray analysis (STEM-EDX) elemental map of the FIB-prepared TEM specimen TKY3 shows that the distribution of the major elements appears to be homogenous (Fig. 3a). Selected area electron diffraction patterns of the particle matrix revealed diffuse diffraction maxima corresponding to the amorphous structure (Fig. 3b); however, the magnified image of the elemental map clearly shows nanoparticles less than ~10 nm in size, which consist mainly of Fe, Zn and Cs. These nanoparticles form distinct domain texture in a Si oxide matrix (Fig. 3c). The Si map reveals mostly homogeneous with slightly deficient contrast at the position of nanoparticles. This is because the volume of nanoparticle is smaller than that of Si oxide matrix considering the thickness of TEM specimen (typically up to

~200 nm). Most of these nanoparticles were identified as exhibiting a franklinite structure ($ZnFe_2O_4$, $Fd3m$, Z = 8)[19] on the basis of the HRTEM image with the fast Fourier transform image (Fig. 3d). Two other TKY CsMPs also revealed the same texture for Zn–Fe-oxide nanoparticles embedded in a $SiO_2$ amorphous glass matrix (Fig. 4a–d). Standardless quantitative analysis of the major constituent elements in TKY3, 4 and 5 yielded results of 54.4–86.6 wt% $SiO_2$, 0.55–10.9 wt% $Cs_2O$, 3.29–13.8 wt% $Fe_2O_3$, 1.48–6.80 wt% $SnO_2$ and 4.33–16.5 wt% ZnO (Table S1), which is similar to the composition reported in the previous study[13]. The OTZ CsMP obtained from Fukushima soil was also composed of $SiO_2$ with Zn, Fe and Cs, with Cs concentrations of 8.50–12.9 wt.% as $Cs_2O$[17]. This particle exhibited the same nanoscale texture as those from Tokyo (Fig. 4e and f). In addition to the major constituents, trace elements such as K, Cl, Sn, Rb, Pb, Mn and Mo were detected by the STEM-EDX area analysis[17].

*Dissolution experiments of the TKY air filters*

To conduct quantitative assessment of the contribution of these CsMPs derived from the FDNPP to total Cs radioactivity, dissolution experiments on TKY air filter #1 were performed in ultra-pure water (Milli-Q) for 24 h at room temperature under static conditions. Autoradiography of a piece of the filter revealed that the position of radioactive spots before the experiment was unchanged after immersion in ultra-pure water, indicating that the CsMPs were insoluble (Fig. S2). The radioactivities of the filter pieces before and after the dissolution experiment were measured as 18.9 and 16.7 Bq of $^{134+137}Cs$, respectively, while 1.61 Bq of $^{134+137}Cs$ was dissolved into the solution (Table 2). The contribution from radioactive Cs bound to clay minerals to the radioactivity of the filter is negligible because the sample was collected from the first atmospheric plume contaminated with radionuclides from the FDNPP, which arrived at 0:00–7:00 on 15th March. In addition, other CsMPs were detected at high-activity spots on the autoradiograph, and most of the remaining radioactivity on the filter was from the CsMPs.

## 4. Discussion

Despite the small particle size and the overall activity being less than a few Bq, the radioactivities per unit mass calculated of the TKY CsMPs are some of the highest activities per unit mass ever encountered in hot particles from nuclear accidents or atomic bomb explosions: for comparison, the activity is 10 to 100 times greater than that of the nuclear fuel particles and comparable to microparticles with beta- and gamma-emitting radionuclides encountered in Chernobyl[20,21]. Consequently, the local radiation field adjacent to CsMPs is very high.

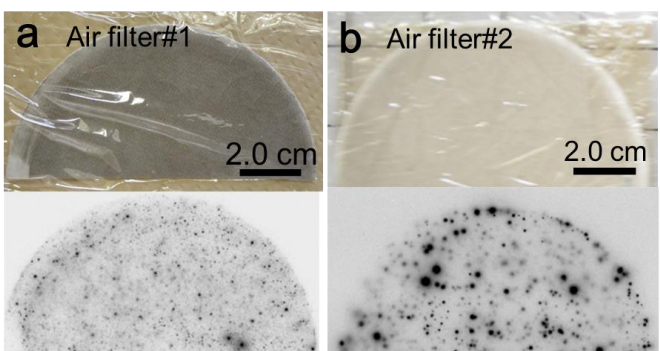

**Figure 1:** (a) and (b) optical image (top) and autoradiograph image (bottom) of air filters #1 and #2, respectively.

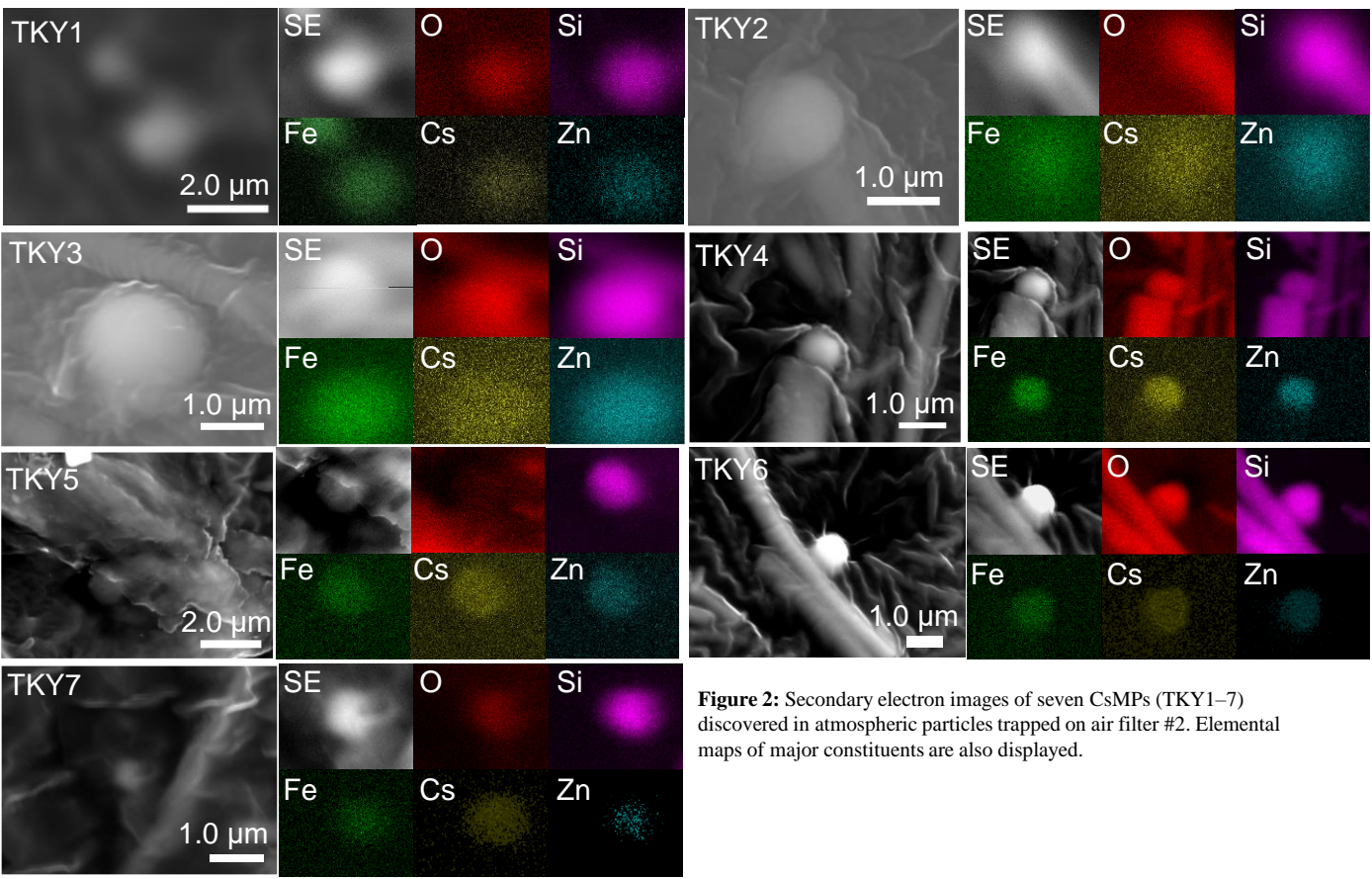

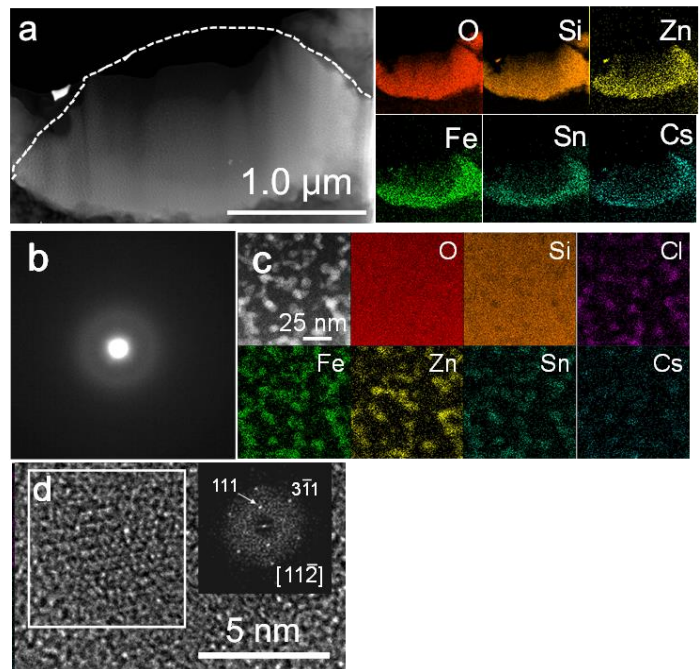

Figure 2: Secondary electron images of seven CsMPs (TKY1–7) discovered in atmospheric particles trapped on air filter #2. Elemental maps of major constituents are also displayed.

The major and trace elements of TKY CsMPs were derived from elements inside the reactor during the meltdowns; however, the compositions are markedly different from those in the debris[22], which consists of a mixture of melted core, reactor materials and concrete. Sn was included in a Zr–Sn alloy; Fe and Mn were constituents of the reactor pressure vessel; Si was derived from siliceous concrete released during molten-core–concrete interaction[17,23,24] and less possibly from insulating materials[14]; Cs, Rb, Pb, Sn and Mo were fission products included in the irradiated fuels; Cl was from seawater, and Zn was generally added to the reactor water to prevent radioactive corrosion of steel by formation of a protective oxide layer. Hence, from the TEM data, the composition and nanotexture of CsMPs from Tokyo (TKY1–7) are the same as those of the particle from Fukushima (OTZ), strongly suggesting that these CsMPs experienced the same formation processes and had an identical origin.

In the Cs fallout in Tokyo arrived at 0:00–7:00 on 15th March, the soluble fraction of Cs species, which may include hydroxides, chloride and iodide, contributed only 11% of the total radioactivity, whereas the contribution of CsMPs to the total Cs radioactivity was as high as 89%. This strongly suggests that CsMPs were responsible for most of the radiation dose derived from radioactive $^{134+137}$Cs (124 Bq/m³)[18] by inhalation and skin contact in addition to the other soluble radionuclides such as $^{131+132}$I (522 Bq/m³)[18] in the initial largest Cs plume to Tokyo, a distance of ~230 km from the FDNPP, on 15th March, 2011.

The CsMPs are significant sources of radioactive Cs to the surrounding environment and the ecosystem. Based on the small size of the particles, 0.58–2.0 μm, which is within the range of PM$_{2.5}$, Cs-rich airborne particles were ubiquitously distributed and readily inhaled deep into the alveolar region of the respiratory system[25]. Almost all of the CsMPs would have passed through the extrathoracic region in the case of oral inhalation, while ~50% of 2-μm-sized particles could have been deposited in that region in the case of nasal inhalation[25,26]. Approximately 20%–50% and <10% of CsMPs could be deposited in the alveolar and tracheobronchial regions, respectively[25,26]. Some of the deposited CsMPs would not be phagocytized by macrophages but instead slowly translocated to lymph nodes, from which the biological half-times of elimination are estimated to be tens of years[27]. Thus, a certain fraction of CsMPs is expected to be retained inside the body compared with the typical half-time of $^{137}$Cs, ~100 days[28].

Figure 3: (a) High-angle annular dark-field scanning transmission electron microscopy (HAADF-STEM) image of the cross-section of TKY3 TEM specimen prepared by focused ion beam (FIB). The white dotted line represents the original shape of the particle before FIB thinning. Elemental maps of the major constituents are also shown. (b) Selected area electron diffraction pattern (c) Enlarged HAADF-STEM image associated with elemental maps. (d) High-resolution transmission electron microscopy image of a Zn–Fe-oxide nanoparticle and the fast Fourier transform image.

**Table 2:** Results of dissolution experiments on two pieces cut from air filter #1, labelled as a and b. The radioactivity was decay-corrected to 00:00 JST on 15th March, 2011.

| Sample | | Radioactivity (Bq) before dissolution experiment | Radioactivity (Bq) after dissolution experiment | Radioactivity (Bq) of filtrate | Recovery of radioactivity (%) | Fraction of residual CsMP in radioactivity (%) |
|---|---|---|---|---|---|---|
| Air filter #1-a | $^{134}$Cs | 9.892(±0.141) | 8.639(±0.088) | 0.768(±0.045) | 95.1 | 87.3 |
|  | $^{137}$Cs | 9.032(±0.057) | 8.018(±0.034) | 0.837(±0.023) | 98.0 | 88.8 |
| Air filter #1-b | $^{134}$Cs | 2.354(±0.090) | 1.901(±0.085) | 0.589(±0.030) | 106 | 80.8 |
|  | $^{137}$Cs | 2.188(±0.032) | 1.870(±0.030) | 0.619(±0.014) | 115 | 85.5 |

The effective dose coefficient that is already present considers the case of inhaling particulate matter as small as 1 μm with slow clearance; however, only the average can be calculated[29]. In case of CsMPs, because of the very high radioactivity per unit mass (~$10^{11}$ Bq/g), the intense β and γ radiation per unit volume will cause radiolysis. In particular, β radiation will produce radical species at a scale of a few hundred microns, which is an order of magnitude greater than the size of macrophages and respiratory epithelial cells[30]. Energy deposition by β radiation within the 100-μm-thick water layer on the surface of the TKY CsMPs was estimated to be 0.51–12 × $10^{-3}$ Gy/h (Table 3). The high-energy deposition can produce various radicals: $H_2$, $H_2O_2$ and H•, with a •OH production rate of as much as 2.5 × $10^3$ molecules/s within the 100 μm-thick water layer on the surface of the CsMPs, which may cause oxidative damage to cellular DNA[31,32].

Even though external exposure and exposure by ingestion have been reported to be some orders of magnitude more important than the dose contribution from inhalation[33], assessment of inhalation exposure has not yet included the impact of CsMPs. The sparingly soluble CsMPs may have longer biological half-times, and hence higher dose contributions, although the overall dose might still be dominated by external exposure. A detailed assessment of the impact of the CsMPs on inhalation dose contributions should be conducted to assess this possibility. Although the contribution of CsMPs in Tokyo atmospheric particles was determined in this study, the contribution of the CsMPs to the total inventory of radioactivity in the area around the FDNPP and surrounding prefectures remains to be determined for accurate assessment of the environmental impact of the Fukushima–Daiichi tragedy.

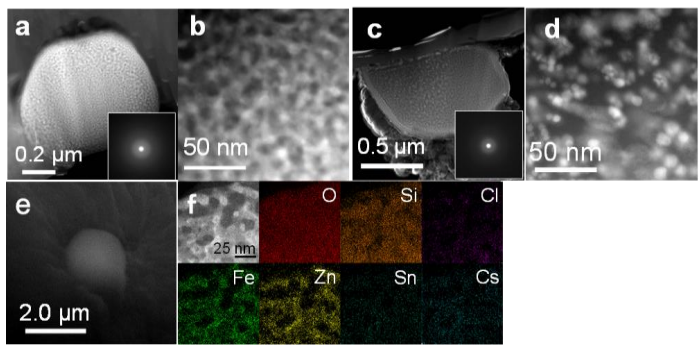

**Figure 4:** (a) HAADF-STEM image of FIB-TEM specimen of TKY4. The inset image shows the SAED pattern. (b) Enlarged HAADF-STEM image of (a). (c) HAADF-STEM image of FIB-TEM specimen of TKY5. The inset image shows the SAED pattern. (d) Enlargement of (c). (e) SEM image of OTZ CsMP. (f) HAADF-STEM image with the elemental maps of major constituents.

## Acknowledgements

The authors are grateful for three anonymous reviewers for their comments and corrections. This study is partially supported by JST Initiatives for Atomic Energy Basic and Generic Strategic Research and by a Grant-in-Aid for Scientific Research (KAKENHI) from the Japan Society for the Promotion of Science (16K12585, 16H04634, No. JP26257402).
SU is also supported by ESPEC Foundation for Global Environment Research and Technology (Charitable Trust) (ESPEC Prize for the Encouragement of Environmental Studies). The authors are grateful to Dr. Watanabe for her assistance on SEM analyses at the Center of Advanced Instrumental Analysis, Kyushu University. The findings and conclusions of the authors of this paper do not necessarily state or reflect those of the JST.

**Table 3:** Summary of the energy deposition by β radiation (Gy/h) and the amounts of chemical species and radicals produced by β ray radiolysis within the 100-μm-thick volume of water over each CsMP. Note that the energy deposition by γ radiation is negligibly small.

| Sample | Energy deposition by β radiation (Gy/h) | molecules/s | | | | |
|---|---|---|---|---|---|---|
| | | $H_2$ | $H_2O_2$ | H• | •OH | •$HO_2$ |
| TKY1 | 1.23 × $10^{-2}$ | 4.15×$10^2$ | 6.43×$10^2$ | 5.47×$10^2$ | 2.47×$10^3$ | 2.38×$10^1$ |
| TKY2 | 4.47 × $10^{-3}$ | 1.50×$10^2$ | 2.33×$10^2$ | 1.98×$10^2$ | 8.96×$10^2$ | 8.66 |
| TKY3 | 5.54 × $10^{-3}$ | 1.88×$10^2$ | 2.92×$10^2$ | 2.48×$10^2$ | 1.12×$10^3$ | 1.08×$10^1$ |
| TKY4 | 2.90 × $10^{-3}$ | 9.72×$10^1$ | 1.51×$10^2$ | 1.28×$10^2$ | 5.77×$10^2$ | 5.57 |
| TKY5 | 2.35 × $10^{-3}$ | 7.90×$10^1$ | 1.23×$10^2$ | 1.04×$10^2$ | 4.69×$10^2$ | 4.53 |
| TKY6 | 2.72 × $10^{-3}$ | 9.07×$10^1$ | 1.41×$10^2$ | 1.20×$10^2$ | 5.42×$10^2$ | 5.22 |
| TKY7 | 5.12 × $10^{-4}$ | 1.70×$10^1$ | 2.64×$10^1$ | 2.24×$10^1$ | 1.01×$10^2$ | 9.77×$10^{-1}$ |
| OTZ | 2.18 × $10^{-2}$ | 7.40×$10^2$ | 1.15×$10^3$ | 9.73×$10^2$ | 4.41×$10^3$ | 4.25×$10^1$ |

# Author Contributions

S.U. conceived the idea, designed all experiments, and wrote the manuscript. S.U. performed measurements and data analysis. G.F. performed dose calculations. A.O. conducted TEM analysis. K.N. provided navigation during field research in Fukushima. S.Y. performed gamma spectroscopy at Tsukuba University. B.G. and R.C.E. participated in the analysis of the data and writing of the manuscript.

Competing financial interests
The authors declare no competing financial interests.

Corresponding authors
Satoshi Utsunomiya

# Supplementary Information

Attached

# Caesium fallout in Tokyo on 15th March, 2011 is dominated by highly radioactive, caesium-rich microparticles

By Satoshi Utsunomiya*, Genki Furuki, Asumi Ochiai, Shinya Yamasaki, Kenji Nanba,

Bernd Grambow, and Rodney C. Ewing

Table S1: Chemical composition (wt.%) of the area analysis on the FIB thin section of the TKY3, 4, 5 and OTZ CsMPs as determined by STEM-EDX standardless quantification. The analysis was conducted using a rastered STEM probe (~100 × 100 nm) to minimize diffusion of alkaline elements such as K and Cs under the electron beam. The composition was calculated as an oxide except for Cl and normalized to 100% in total. Iron concentration is presented as $Fe_2O_3$.

|  | $SiO_2$ | $Fe_2O_3$ | ZnO | $Cs_2O$ | $SnO_2$ | $Rb_2O$ | $K_2O$ | Cl | MnO | PbO |
|---|---|---|---|---|---|---|---|---|---|---|
| *TKY3* | | | | | | | | | | |
| Area 1 | 82.0 | 5.94 | 4.81 | 1.59 | 2.92 | 0.60 | 1.28 | 0.33 | 0.27 | 0.30 |
| Area 2 | 81.8 | 6.51 | 4.33 | 2.09 | 2.44 | 0.00 | 2.04 | 0.32 | 0.49 | 0.00 |
| Area 3 | 79.6 | 7.61 | 5.05 | 1.66 | 2.53 | 0.91 | 1.21 | 0.25 | 0.50 | 0.69 |
| Area 4 | 85.2 | 5.04 | 4.75 | 0.81 | 2.75 | 0.00 | 0.04 | 1.18 | 0.24 | 0.00 |
| Area 5 | 84.7 | 5.09 | 4.48 | 0.99 | 3.48 | 0.00 | 0.04 | 1.21 | 0.05 | 0.00 |
| Area 6 | 81.0 | 7.48 | 5.11 | 1.69 | 2.76 | 0.41 | 0.94 | 0.31 | 0.32 | 0.00 |
| Area 7 | 77.8 | 7.86 | 6.05 | 1.70 | 3.27 | 0.95 | 0.88 | 0.29 | 0.48 | 0.73 |
| Area 8 | 82.2 | 6.82 | 5.3 | 1.77 | 2.68 | 0.00 | 0.54 | 0.40 | 0.33 | 0.00 |
| Area 9 | 79.7 | 7.17 | 6.14 | 1.71 | 3.04 | 0.51 | 0.88 | 0.32 | 0.23 | 0.30 |
| Area 10 | 83.9 | 5.59 | 5.15 | 1.03 | 3.06 | 0.00 | 0.10 | 0.79 | 0.37 | 0.00 |
| Area 11 | 80.9 | 6.71 | 5.17 | 1.57 | 2.68 | 0.71 | 1.34 | 0.26 | 0.34 | 0.30 |
| Area 12 | 81.5 | 7.10 | 5.90 | 1.41 | 2.89 | 0.00 | 0.55 | 0.30 | 0.310 | 0.00 |
| Area 13 | 84.1 | 5.74 | 5.11 | 1.23 | 2.55 | 0.00 | 0.23 | 0.64 | 0.40 | 0.00 |
| Area 14 | 81.4 | 7.60 | 5.43 | 1.56 | 2.24 | 0.00 | 0.98 | 0.38 | 0.44 | 0.00 |
| Area 15 | 82.5 | 7.21 | 5.16 | 0.85 | 2.66 | 0.00 | 0.62 | 0.49 | 0.54 | 0.00 |
| Area 16 | 79.0 | 7.83 | 5.5 | 1.94 | 3.09 | 0.00 | 1.97 | 0.37 | 0.31 | 0.00 |
| Area 17 | 80.0 | 7.17 | 5.25 | 1.87 | 3.09 | 0.00 | 1.94 | 0.20 | 0.49 | 0.00 |
| *TKY4* | | | | | | | | | | |
| Area 18 | 55.7 | 12.6 | 14.2 | 9.85 | 3.80 | 1.02 | 0.83 | 1.29 | 0.61 | 0.07 |

| | | | | | | | | | |
|---|---|---|---|---|---|---|---|---|---|
| Area 19 | 55.0 | 11.9 | 14.5 | 10.8 | 3.49 | 1.20 | 0.90 | 1.33 | 0.45 | 0.47 |
| Area 20 | 54.4 | 12.3 | 15.6 | 10.9 | 1.72 | 1.45 | 1.11 | 1.42 | 0.42 | 0.75 |
| Area 21 | 56.7 | 11.6 | 14.2 | 10.3 | 2.90 | 0.95 | 1.11 | 1.25 | 0.45 | 0.57 |
| Area 22 | 59.3 | 13.7 | 15.0 | 7.27 | 1.97 | 0.42 | 0.46 | 1.33 | 0.36 | 0.23 |
| Area 23 | 56.8 | 13.6 | 16.1 | 8.16 | 1.51 | 0.49 | 1.21 | 1.35 | 0.36 | 0.41 |
| Area 24 | 57.5 | 13.4 | 16.5 | 7.58 | 1.48 | 0.41 | 1.00 | 1.24 | 0.37 | 0.53 |
| Area 25 | 58.9 | 13.3 | 16.3 | 5.84 | 1.84 | 0.39 | 0.95 | 1.59 | 0.50 | 0.42 |
| Area 26 | 56.8 | 13.8 | 16.1 | 7.30 | 2.41 | 0.54 | 0.68 | 1.59 | 0.47 | 0.39 |
| Area 27 | 54.8 | 13.7 | 16.3 | 9.55 | 1.75 | 0.87 | 1.19 | 1.39 | 0.50 | 0.00 |
| TKY5 | | | | | | | | | | |
| Area 28 | 82.1 | 3.64 | 5.64 | 1.47 | 3.02 | 1.97 | 1.59 | 0.26 | 0.00 | 0.34 |
| Area 29 | 67.1 | 7.00 | 6.82 | 8.18 | 6.51 | 2.95 | 0.49 | 0.74 | 0.21 | 0.00 |
| Area 30 | 83.3 | 3.41 | 5.37 | 1.95 | 2.73 | 1.88 | 1.04 | 0.24 | 0.04 | 0.02 |
| Area 31 | 70.8 | 7.52 | 6.20 | 6.61 | 6.80 | 0.62 | 0.43 | 0.70 | 0.10 | 0.25 |
| Area 32 | 83.0 | 3.29 | 5.31 | 1.88 | 3.04 | 1.89 | 1.18 | 0.30 | 0.08 | 0.00 |
| Area 33 | 82.6 | 3.50 | 5.61 | 1.56 | 3.18 | 2.02 | 1.16 | 0.29 | 0.05 | 0.02 |
| Area 34 | 84.2 | 3.48 | 5.63 | 1.05 | 2.84 | 1.42 | 0.91 | 0.22 | 0.22 | 0.01 |
| Area 35 | 81.6 | 3.99 | 5.96 | 1.79 | 2.91 | 1.09 | 2.16 | 0.24 | 0.02 | 0.26 |
| Area 36 | 86.6 | 3.65 | 4.94 | 0.55 | 3.15 | 0.38 | 0.31 | 0.15 | 0.01 | 0.21 |
| Area 37 | 80.8 | 4.36 | 6.28 | 1.78 | 3.56 | 1.03 | 1.60 | 0.29 | 0.09 | 0.17 |
| Area 38 | 80.8 | 4.52 | 5.61 | 3.41 | 3.13 | 0.82 | 1.22 | 0.25 | 0.00 | 0.27 |
| Area 39 | 83.3 | 4.16 | 6.01 | 1.16 | 3.01 | 0.72 | 1.18 | 0.26 | 0.03 | 0.16 |
| Area 40 | 81.4 | 4.05 | 6.40 | 1.40 | 3.41 | 0.94 | 1.93 | 0.22 | 0.14 | 0.10 |
| Area 41 | 80.1 | 4.58 | 5.72 | 3.66 | 3.27 | 0.71 | 1.55 | 0.26 | 0.14 | 0.01 |
| Area 42 | 81.2 | 4.09 | 5.77 | 2.28 | 2.75 | 1.06 | 2.01 | 0.24 | 0.07 | 0.50 |

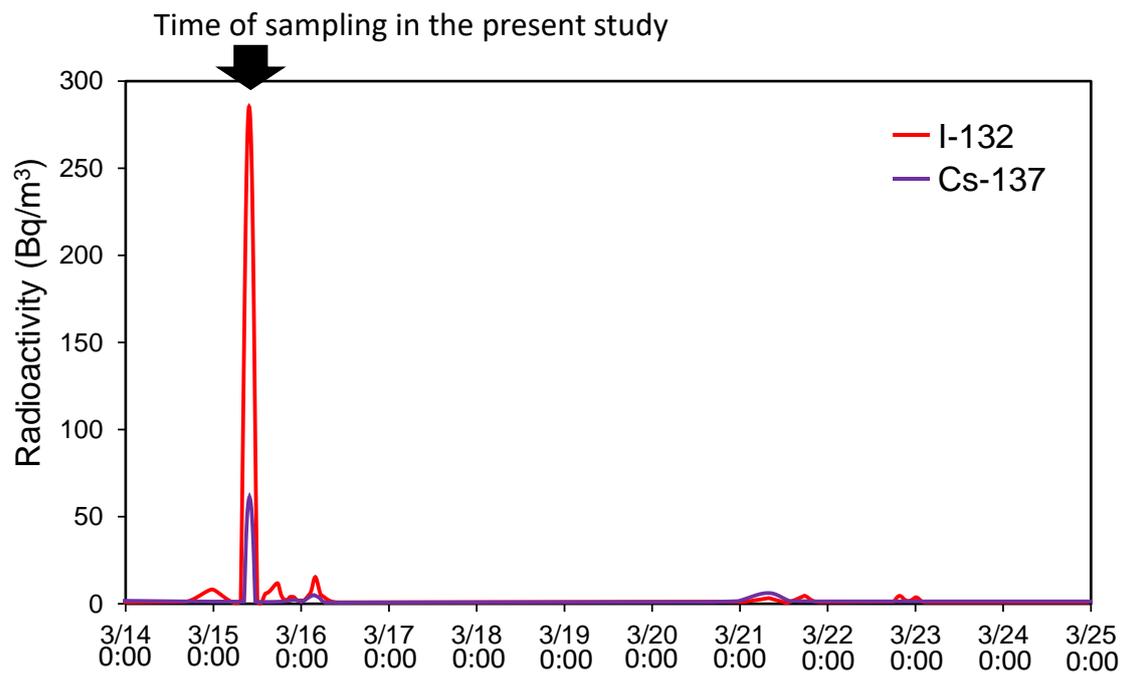

Figure S1. Transition of radioactivities of two selected radionuclides, $^{132}$I and $^{137}$Cs, detected at Setagaya ward, Tokyo during 14-25 of March, 2011. Modified after Nagakawa et al. (2011)[18].

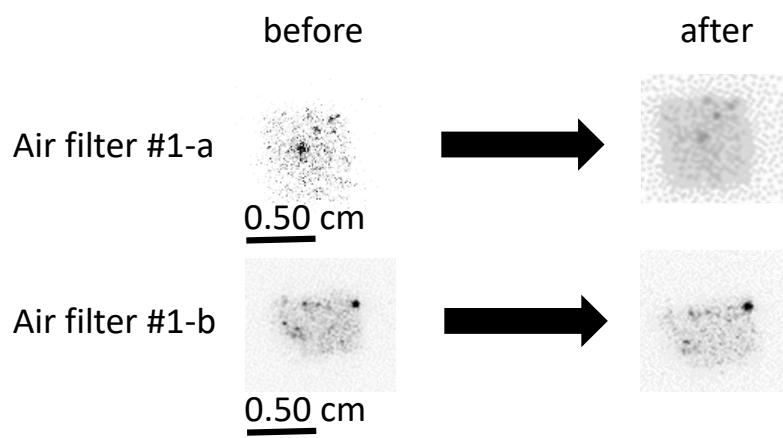

Figure S2: Autoradiographs of two pieces of air filter #1 from Tokyo before and after the dissolution experiments wherein the filters were immersed in ultra-pure water for 24 h at room temperature.

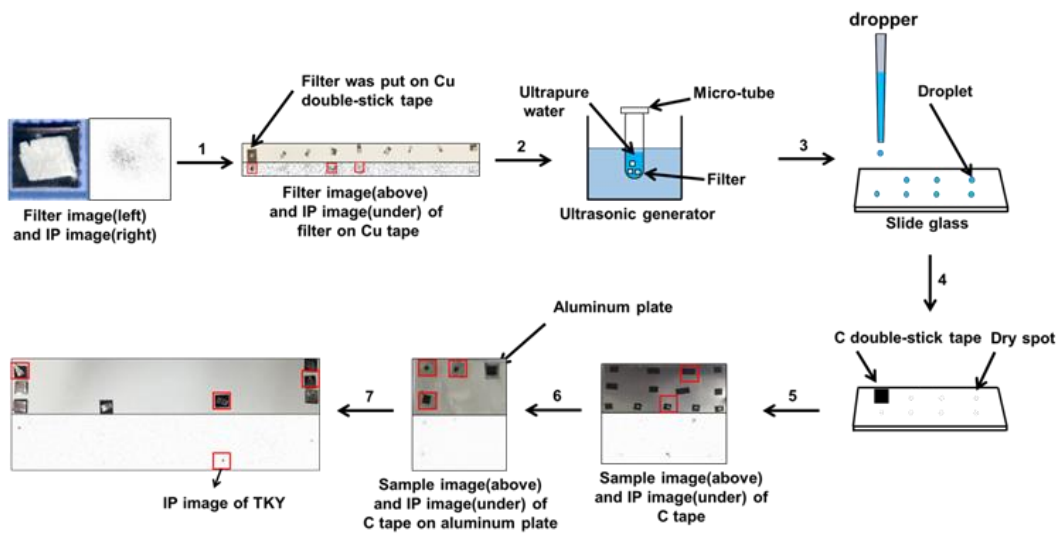

Figure S3: Diagram showing the procedure used to separate Cs-rich microparticles from an air filter. (1) The filter was cut into nine pieces. (2) Pieces of filter with strongly radioactive spots were suspended in ultra-pure water and ultra-sonicated. (3) Drops of the suspension were placed on a glass slide. (4) The droplets were allowed to dry completely. (5) The dried portion of the droplets was attached with double-stick carbon tape and the pieces of the tape were investigated by IP. (6) The pieces were cut into pieces that were as small as possible and placed on an aluminium plate for further analysis.

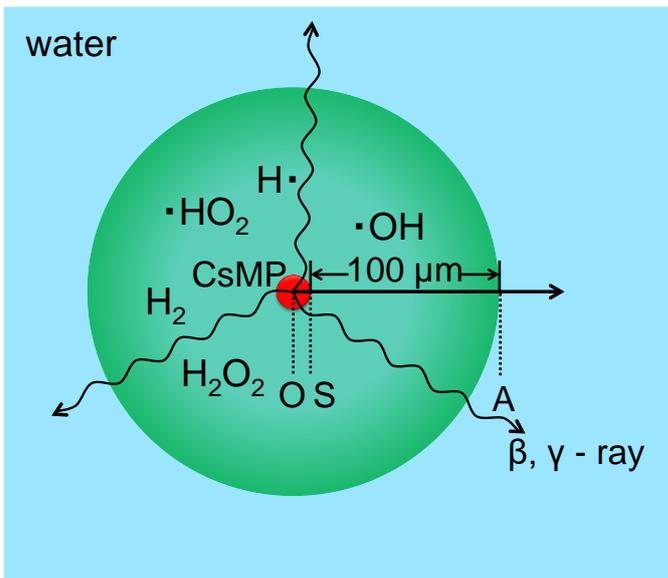

Figure S4: Schematic diagram illustrating radiolysis around a CsMP.